\documentclass[a4paper]{aa}
\usepackage{graphicx,float}

\def \sax {BeppoSAX}
\def \src {XB\thinspace1832$-$330}
\def \ngc {NGC\thinspace6652}

\def \hcm {\hbox {\ifmmode $ atom cm$^{-2}\else atom cm$^{-2}$\fi}}
\def \arcmin {\hbox{$^\prime$}}

\def\approxgt{\mathrel{\hbox{\rlap{\lower.55ex \hbox {$\sim$}}
        \kern-.3em \raise.4ex \hbox{$>$}}}}
\def\approxlt{\mathrel{\hbox{\rlap{\lower.55ex \hbox {$\sim$}}
        \kern-.3em \raise.4ex \hbox{$<$}}}}

\begin{document}

\title{BeppoSAX study of the X-ray binary 
\src\ located in the globular cluster \ngc}

\author{A.N. Parmar\inst{1}
        \and T. Oosterbroek\inst{1} 
        \and L. Sidoli\inst{1}
        \and L. Stella\inst{2}
        \and F. Frontera\inst{3}
}
\offprints{A.N.Parmar (aparmar@astro.estec. esa.nl)}

\institute{
       Astrophysics Division, Space Science Department of ESA, ESTEC,
       Postbus 299, NL-2200 AG Noordwijk, The Netherlands
\and  
       Osservatorio Astronomico di Roma, Via Frascati 33,
       Monteporzio Catone, I-00040 Roma, Italy
\and
       Istituto Tecnologie e Studio Radiazioni Extraterrestri, CNR, 
       Via Gobetti 101, I-40129 Bologna, Italy
}

\date{Received 22 August 2001 / Accepted}

\titlerunning{BeppoSAX observation of \src}

\abstract{
Results of a 2001 March 17--20 BeppoSAX observation 
of the X-ray binary
\src\ located in the globular cluster \ngc\ are presented.
In contrast to the majority of luminous globular cluster X-ray sources, 
the (0.1--200~keV)
BeppoSAX spectrum cannot be fit with a disk-blackbody and
{\sc comptt} Comptonization model, unless partial covering
is included. This confirms the ASCA detection of partial covering by
Mukai \& Smale~(\cite{m:00}).
The best-fit spectral parameters are similar to those
of the globular cluster sources with orbital periods of $<$1~hr,
implying that \src\ is also an ultra-compact system.
A plausible optical candidate to \src\
has recently been discovered by Heinke et al.~(\cite{h:01})
which shows evidence for possible 0.92, 2.22, or 4.44~hr periodicities.
We find no evidence for any 2--10~keV periodic modulation
at any of these periods with a
2$\sigma$ upper limit to the semi-amplitude of any sinusoid 
of $\approxlt$1.5\%. 
}
\maketitle

\section{Introduction}
\label{sect:intro}

The X-ray source \src\ in NGC\thinspace6652 is one of 12 bright 
($L{\rm _x \approxgt 10^{36}}$~erg~s$^{-1}$) low-mass 
X-ray binaries (LMXRBs) located
in galactic globular clusters. 
Although the uncertainty region for the HEAO-1 source H\thinspace1825-331
contained \ngc, it was not originally considered to be a secure
identification because of its 2.7 deg$^2$ area (Hertz \& Wood \cite{h:85}).
A secure identification was obtained during the
ROSAT All-Sky Survey when a bright
source was detected co-incident with
\ngc\ (Predehl et al. \cite{phv:91}). The source was subsequently
detected in pointed ROSAT observations (Johnston et al.~\cite{j:96}) and
Type I X-ray bursts and persistent emission were observed
by the BeppoSAX Wide Field Cameras (in 't Zand et al. \cite{i:98})
confirming that the compact object in the system is
a neutron star.
From HST observations, a relatively low luminosity star (M${\rm _V}$ = +4.7),
located 2.3$\sigma$ from the ROSAT position, which
exhibits a $43.6 \pm 0.6$~minute modulation
was proposed as the counterpart by Deutsch et al. (\cite{dma:98},
\cite{d:00}).
However, recent {\it Chandra} High Resolution Camera-Imager (HRC-I)
observations revealed three additional
faint X-ray sources in \ngc\ and the optical counterpart proposed by
Deutsch et al. (\cite{dma:98}, \cite{d:00}) is clearly associated with 
one of these (Heinke et al.~\cite{h:01}). Instead, \src\ is
identified with a blue variable object with M${\rm _V}$ = +3.7
(Heinke et al.~\cite{h:01}). Possible periods of 0.92, 2.22 or 4.44~hours
as well as non-periodic flickering were observed.
This absolute V band magnitude is comparable
with those of ultra-compact LMXRBs in globular clusters (+3.7 for
X\thinspace1820-303 in NGC\thinspace6624 and +5.6 for X\thinspace1850-087
in NGC\thinspace6712; van Paradijs \& McClintock \cite{pc:94}), although
a low-mass M dwarf companion cannot be excluded.

As part of a program to observe diffuse
galactic emission \src\ was serendipitously 
observed by ASCA (Mukai \& Smale~\cite{m:00}). Although the set-up was
far from ideal (no GIS2 instrument, 4-CCD mode for both SIS instruments
with the counts spread
over all 4 CCD devices), a 0.7--10~keV spectrum was obtained.
Mukai \& Smale (\cite{m:00}) had difficulty in finding an acceptable spectral
model and used partial covering of a power-law continuum  
as their best-fit model. However, as they comment, given the
ASCA calibration uncertainties and the narrow energy range of
the data, it is not possible to say whether this spectral shape is unique, 
or preferred. We note that the {\it Chandra} observation of 
Heinke et al.~(\cite{h:01}) used the HRC-I which does not provide spectral
information. 

We report on a BeppoSAX observation of \src\ made to complete the
systematic study of the persistent luminous globular cluster X-ray sources
reported in Sidoli et al.~(\cite{s:01}) and references within. 
\ngc\ is a compact globular cluster located $\approx$2~kpc from the Galactic
center (Chaboyer et al.~\cite{c:00}). 
It has a low reddening ($E{\rm _{B-V} = 0.10 \pm 0.02}$) and an 
abundance [Fe/H] $\approx$ $-$0.9 (Ortolani et al.~\cite{o:94}). 
This redenning corresponds to an absorption,   
$N{\rm _H}$, of $\sim$$7 \, 10^{20}$~atom~cm$^{-2}$, 
using the relation between
$A{\rm _v}$ and $N{\rm _H}$ of Predehl \& Schmitt (\cite{ps:95}).

\section{Observations}
\label{sect:obs}

Results from the Low-Energy Concentrator Spectrometer (LECS;
0.1--10~keV; Parmar et al. \cite{p:97}), the Medium-Energy Concentrator
Spectrometer (MECS; 1.8--10~keV; Boella et al. \cite{b:97}),
and the Phoswich
Detection System (PDS; 15--300~keV; Frontera et al. \cite{f:97}) on-board \sax\
are presented. 
The MECS consists of two grazing incidence
telescopes with imaging gas scintillation proportional counters in
their focal planes. The LECS uses an identical concentrator system as
the MECS, but with an ultra-thin entrance window
to extend the low-energy response to 0.1~keV.
The non-imaging PDS consists of four units arranged in pairs 
each having a
separate collimator. 

The region of sky containing \src\ was observed by \sax\
on 2001 March 17 19:39 to March 20 08:59 UT.
Good data were selected in the usual way
using the SAXDAS 2.1.1 data analysis package.
A PDS collimator dwell time of 96~s for each on- and
off-source position was used with a rocking angle
of 210\arcmin.
LECS and MECS data were extracted centered on the position of \src\ 
using radii of 8\arcmin\ and 4\arcmin, respectively.
The exposures 
in the LECS, MECS, and PDS are 20.3~ks, 53.3~ks,
and 29.3~ks, respectively. 
Background subtractions for the LECS and MECS
were performed using standard files, but is not critical for such a
bright source. 
Background subtraction for the 
PDS was performed using data
obtained during intervals when the collimators were offset
from the source. 

\begin{figure}
 \centerline{\includegraphics[width=5.0cm,angle=-90]{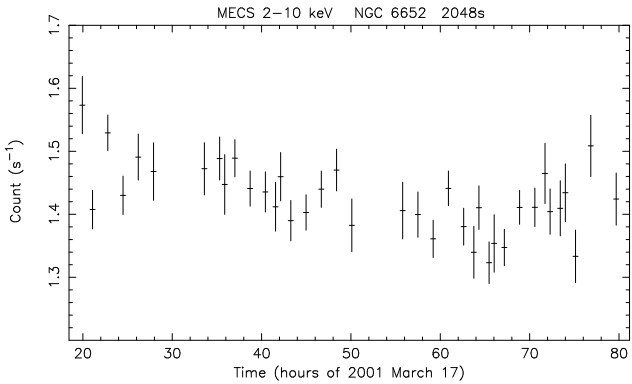}}
  \caption[]{The MECS 2--10 keV lightcurve of \src\ plotted with a time
  resolution of 2048~s}
  \label{fig:lightcurve}
\end{figure}

\section{Results}

\subsection{Lightcurve}
\label{subsect:lightcurve}

Fig.~\ref{fig:lightcurve} shows the 2--10~keV 
MECS lightcurve of \src\ with a binning of 2048~s.
There is no evidence for any X-ray bursts.
The source exhibits variability, in excess of that predicted
from counting statistics, with a root mean square
variability of $2.7 \pm 0.7$\% on a timescale of 2048~s.
Searches for variability around the 0.92, 2.22 and 4.44~hour
possible periods of Heinke et al.~(\cite{h:01}) and the 
0.73~hour period of Deutsch et al. (\cite{d:00}) 
were performed on a 2--10~keV MECS lightcurve with a 256~s binning
using period-folding and a Lomb-Scargle periodogram.
We find no
evidence for significant signals near any of the above
periods. 
The data were folded at each of the proposed periods and
a folded profile produced using 8 phase bins. Unsurprisingly, no
significant evidence for a modulation was present.
The profiles were then fit with a sine function with a fixed period.
The 2$\sigma$ upper limits to the semi-amplitudes
of any such modulations  
are 0.8\%, 1.4\% and 1.4\% for the 0.92, 2.22 and 4.44~hr periods, 
respectively.

\subsection{Spectrum}
\label{subsect:spectrum}

The overall spectrum of \src\ was first investigated by simultaneously
fitting data from the LECS, MECS and PDS using
{\sc xspec} version 11. We note that the three faint X-ray
sources detected in \ngc\ by Heinke et al.~(\cite{h:01}) 
using the {\it Chandra} HRC-I each
have X-ray luminosities $\sim$100 times lower than \src, and are unlikely
to contribute significantly to the observed counts.
The LECS and MECS spectra were rebinned to oversample the full
width half-maximum of the energy resolution by
a factor 3 and to have additionally a minimum of 20 counts 
per bin to allow use of the $\chi^2$ statistic. 
The PDS 
spectrum was rebinned using standard techniques in SAXDAS.
1\% uncertainties were included to allow for systematic uncertainties
in the instrumental responses. All spectral uncertainties
are given at 90\% confidence.
Data were selected in the energy ranges
0.1--4.0~keV (LECS), 1.8--10~keV (MECS),
and 15--200~keV (PDS), 
where the instrument responses are well determined and sufficient
counts obtained. 
This gives 
background-subtracted count rates of 0.69, 1.45, and 1.74~s$^{-1}$ 
for the LECS, MECS, and PDS, respectively. 
The photoelectric absorption
cross sections of Morrison \& McCammon (\cite{m:83}) 
are used throughout.
We note that \src\ 
is detected at higher
energies than any of the other
globular cluster LMXRBs observed with BeppoSAX (see Sidoli et al.~\cite{s:01}).

Initially, simple models were tried, including absorbed power-law,
thermal bremsstrahlung and a cutoff power-law 
($E^{-\alpha}\exp-(E_{{\rm c}}/kT)$).
Factors were included in the spectral fitting to allow for normalization 
uncertainties between the instruments. These factors were constrained
to be within their usual ranges during the fitting. 
A power-law with a photon index, $\alpha$, = 1.8 and low-energy absorption 
equivalent to $4 \, 10^{21}$~atom~cm$^{-2}$ and a 
18~keV bremsstrahlung  
give unacceptable fits with $\chi ^2$'s of 510 and 1127 for 255 
degrees of freedom (dof), respectively.

Sidoli et al.~(\cite{s:01}) conducted a spectral survey of the
luminous globular cluster X-ray sources using BeppoSAX. They found that
with the exception of the accretion disk corona source X\thinspace2127+119
(NGC\thinspace7078) and at times
X\thinspace1820-303 (NGC\thinspace6624), all the other globular 
cluster LMXRB spectra may be acceptably fit using a multicolor
disk-blackbody (Mitsuda et al. \cite{m:84};  
Makishima et al. \cite{m:86}) where $r{\rm _{in}}$ is the innermost
radius of the disk, $i$ the disk inclination angle and
$kT{\rm _{in}}$ the blackbody effective temperature at $r{\rm _{in}}$, 
and the {\sc comptt} Comptonization
model (Titarchuk~\cite{t:94}) where $kT{\rm _e}$ is the temperature 
of the Comptonizing electrons, $\tau_{\rm e}$ the plasma optical depth 
with respect to electron scattering and $kT{\rm _W}$ the temperature
of the input photon (Wien) distribution.
Surprisingly, this model does
not provide an acceptable fit to the \src\ spectrum, giving a 
$\chi ^2$ of 532.2 for 252 dof.

\begin{figure*}
  \centerline{
  \hbox{\hspace{0.0cm}
   \includegraphics[width=6.5cm,angle=-90]{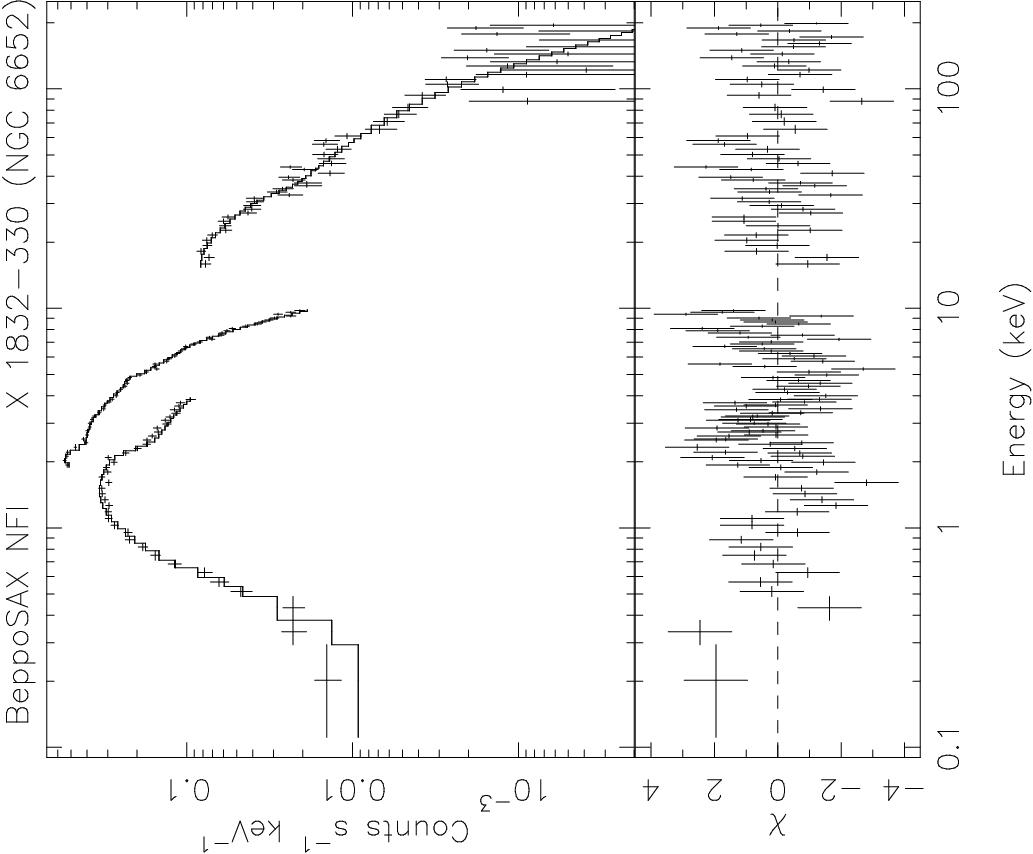}
    \hspace{0.5cm}   
    \includegraphics[width=6.5cm,angle=-90]{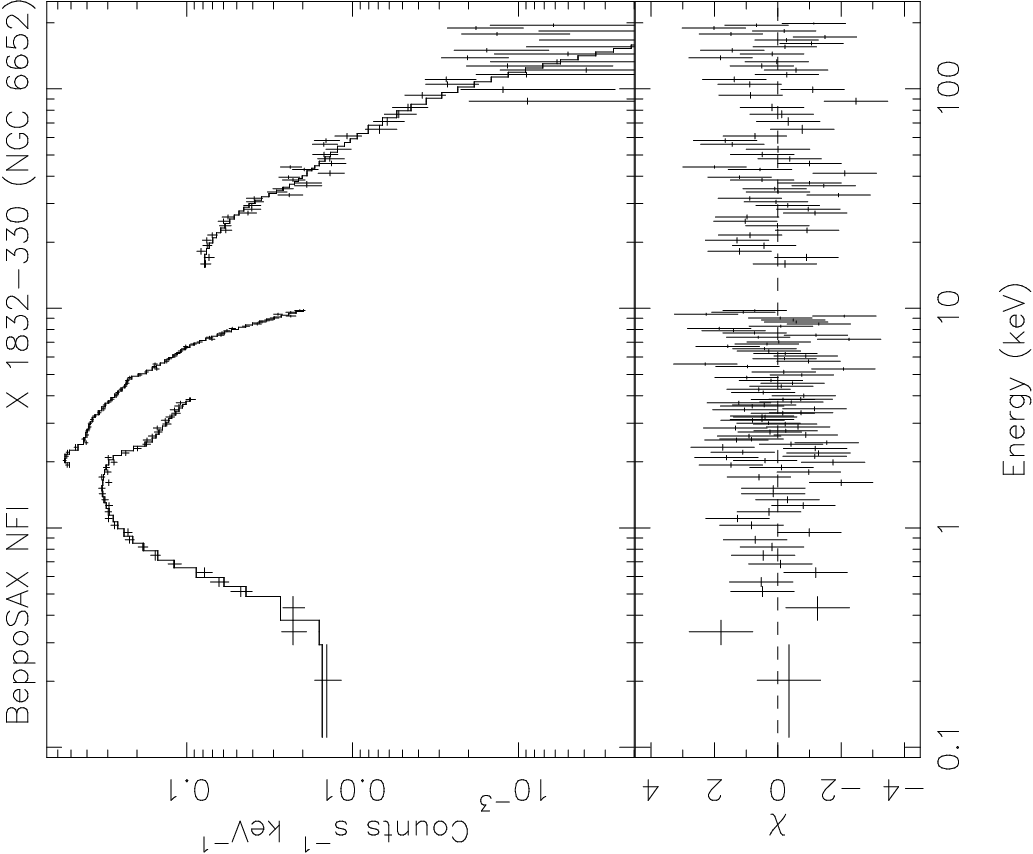}
   \hspace{0.5cm}
   \includegraphics[width=6.5cm,angle=-90]{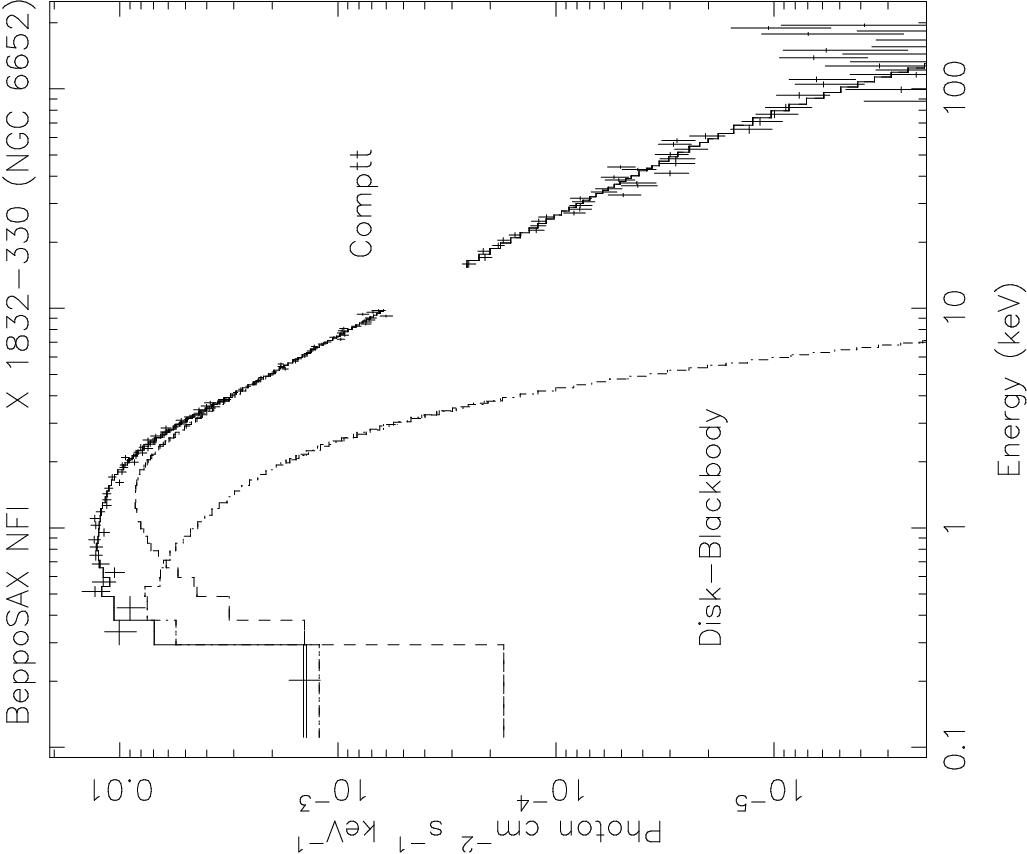}}}
  \caption[]{The overall \src\ NFI spectrum together with the best-fit 
             partially covered cutoff power-law (left) or 
             disk-blackbody and {\sc comptt} (center) model fits 
             (see Table~\ref{tab:spec_paras}). 
             The lower panels show the contributions to $\chi ^2$
             for the two models with the same scaling. 
             The right panel shows the assumed photon spectrum with
             the contributions of the disk-blackbody and {\sc comptt}
             components indicated separately}
  \label{fig:spectrum}
\end{figure*}

The same model as used by Mukai \& Smale (\cite{m:00}) 
in the energy range 0.7--10~keV consisting
of a partially absorbed power-law gives a significantly
better fit with a $\chi ^2$ of 364.3 for 253 dof, suggesting that partial
covering may be important. Examination of the residuals
shows that this model does not well account for an overall curvature
in the spectrum and the power-law was replaced with a cutoff 
power-law to give a $\chi ^2$ of 334.5 for 252 dof. The decrease in
$\chi ^2$ corresponds to an 
F-statistic of 22.5. The probability of such
a decrease occurring by chance is very small ($3.6 \, 10^{-6}$).
We note that the cutoff energy of $200 \, \pm \, ^{130} _{45}$~keV
is well above the upper energy of ASCA. 
This model accounts for the {\it overall} shape of the
0.1--200~keV spectrum {\it reasonably} well 
(Fig.~\ref{fig:spectrum}, left panel). The
best-fit parameters are given in Table~\ref{tab:spec_paras}.

Examination of Fig.~\ref{fig:spectrum} shows that there are still
structured residuals remaining, especially between 2--3~keV
and above 6~keV. 
The size of these features
($\sim$10\%) is larger then the calibration uncertainties of the
BeppoSAX instruments.
Sidoli et al.~(\cite{s:01}) demonstrate that the BeppoSAX spectra
of most luminous globular cluster LMXRBs can be modeled using
a combination of a disk-blackbody and a {\sc comptt} Comptonized 
continuum, while the results of Mukai \& Smale~(\cite{m:00}) (and see
above) indicate that partial covering may be important. 
We therefore investigated whether partial covering of the
standard Sidoli et al.~(\cite{s:01}) spectral model 
could produce the observed spectral shape. A fit using a partially
covered disk-blackbody and {\sc comptt}
Comptonized component gives a $\chi ^2$ of 286.8 for 249 dof.
The best-fit parameters are given in Table~\ref{tab:spec_paras} and
the spectrum shown in Fig.~\ref{fig:spectrum}. The residuals are
significantly less structured.
The decrease in $\chi ^2$ compared to the partially covered cutoff power-law
corresponds to an F-statistic of 41.1. The probability of such
a decrease occurring by chance is $<$$10^{-20}$. 
We thus conclude that the intrinsic X-ray spectrum of \src\
may be modeled in the same way as the majority of other globular cluster 
LMXRB, as long as partial covering is included.
The $N {\rm _H}$ of 
$(4.6 \, \pm \, _{1.2} ^{2.3}) \, 10^{20}$~atom cm$^{-2}$ is
consistent with that predicted from the redenning to \ngc\
of $7 \, 10^{20}$~atom~cm$^{-2}$.
The absorption (both interstellar and partial covering)
corrected 0.1--100~keV ratio of
disk-blackbody to {\sc comptt} fluxes is 0.07. 
The 95\% confidence upper limit equivalent width to a narrow
emission line at 6.4~keV is 50~eV. In order
to investigate whether reflection from cold material with
cosmic abundances could be important, we replaced the
partial covering
by the {\sc pexrav} model in {\sc xspec} and re-fitted the spectrum.
We were unable to obtain an acceptable fit with a 
$\chi ^2$ of 354.7 for 247 dof.

\begin{table}
\begin{center}
\caption[]{Fits to the 0.1--200~keV BeppoSAX spectrum of \src\
using a partially covered cutoff power-law model (upper) and a
partially covered disk-blackbody and {\sc comptt} model (lower).
$f{\rm _{PCF}}$ is the fraction of the flux that undergoes extra absorption, 
$N{\rm _{PCF}}$. 
$L$ is the luminosity in erg~s$^{-1}$ 
for a distance of 9.2~kpc (Chaboyer et al.~\cite{c:00})}
\begin{tabular}{ll}
\hline
\noalign {\smallskip}
Parameter              &  Value \\
\hline
\noalign {\smallskip}
$N {\rm _H}$ ($10^{22}$~atom cm$^{-2}$) & $0.18 \, \pm \, _{0.10} ^{0.05}$ \\

$f{\rm _{PCF}}$                   & $0.67 \, \pm \, ^{0.09} _{0.03}$ \\

$N{\rm _{PCF}}$ ($10^{22}$ atom cm$^{-2}$) 
& $0.83 \, \pm \, _{0.06} ^{0.07}$ \\

$\alpha$           & $1.90 \,  \pm \, _{0.02} ^{0.03}$ \\

$E{\rm _{c}}$~(keV) & $ 200 \, \pm \, ^{130} _{45}$ \\

$\chi ^2$/dof            & 334.5/252 \\

\noalign {\smallskip}
\hline
\noalign {\smallskip}

$N{\rm _H}$  ($10^{20}$~atom cm$^{-2}$) & $4.6 \, \pm \, _{1.2} ^{2.3}$ \\

$f{\rm _{PCF}}$                       & $0.41 \pm  0.05$ \\

$N{\rm _{PCF}}$ ($10^{22}$ atom cm$^{-2}$) 
& $1.70 \, \pm \, _{0.25} ^{0.32}$ \\

$kT{\rm _{in}}$ (keV)     & $0.64 \pm 0.04$ \\
$r_{\rm in} (\cos \theta)^{0.5}$ (km)& 
$2.9 \,  \pm \, ^{0.8} _{0.4}$   \\

$kT{\rm _W}$  (keV)      & $0.41 \pm 0.02$ \\
$kT{\rm _e}$ (keV)       & $25.3 \pm 1.8$ \\

${\rm \tau_p}$           & $1.77  \pm 0.07$ \\

$L$ (1.0--10 keV)    &   $1.6 \, 10^{36}$ \\
$L$ (0.1--200 keV)    &   $4.4 \, 10^{36}$ \\

$\chi ^2$/dof            & 286.8/249 \\

\noalign {\smallskip}                       
\hline
\label{tab:spec_paras}
\end{tabular}
\end{center}
\end{table}

\section{Discussion}
\label{sect:discussion}

Results of a BeppoSAX observation 
of the X-ray binary
\src\ located in the globular cluster \ngc\ are presented.
Sidoli et al.~(\cite{s:01}) showed that the BeppoSAX spectra
of most luminous globular cluster LMXRBs can be modeled using
a combination of a disk-blackbody and a {\sc comptt} Comptonized 
continuum. However, this model does not provide an 
acceptable fit to the \src\ spectrum, unless partial covering
is included. This confirms the result of 
Mukai \& Smale~(\cite{m:00}) who found that
partial covering was required
to obtain an acceptable fit to an ASCA spectrum of \src.
However, the above partially covered 
two-component model gives a significantly
better fit to the BeppoSAX spectrum
than the partially covered power-law (or cutoff power-law)
used by Mukai \& Smale~(\cite{m:00}).

Sidoli et al.~(\cite{s:01}) found that the luminous globular cluster
sources can be divided
into two groups. In the first group, consisting of the 3 known
ultra-compact (orbital period $<$1~hr) sources, the disk-blackbody
temperatures and inner-radii appear physically realistic and the
Comptonization seed photons temperatures and radii of the emission
areas are consistent with the disk temperatures and inner radii.
For all the other 7 sources studied, the disk-blackbody temperatures
appear to be physically unrealistic and the Comptonization parameters
are unrelated to those of the disk-blackbody emission.
For \src\ the $kT{\rm _{in}}$ of $0.64 \pm 0.04$~keV
is similar to those of the ultra-compact systems in NGC\thinspace1851,
NGC\thinspace6712 and NGC\thinspace6624 all of which have $kT{\rm _{in}
\approxlt 1}$~keV, whereas the other sources have $kT{\rm _{in}}$ 
between 1.5--3.5~keV (Sidoli et al.~\cite{s:01}). The
projected inner disk radius, $r{\rm _{in}}$ 
of $2.9 \, \pm \, ^{0.8} _{0.4}$~km
is also closer to those of the above ultra-compact sources
which all have $r{\rm _{in} \approxgt 4}$~km, whereas the other sources
have  $r{\rm _{in} \approxlt 1}$~km. Additionally, as with the
above ultra-compact sources, the temperature of the Comptonization
seed photons, $kT{\rm _W}$, of $0.41 \pm 0.02$~keV is comparable to
$kT{\rm _{in}}$ whereas in the other sources it is a factor $\sim$5 lower.
These comparisons lead us to propose that \src\ is also an ultra-compact
system.

Following a precise {\it Chandra} position for
\src, a plausible optical candidate 
has recently been proposed by Heinke et al.~(\cite{h:01}).
The candidate is variable with evidence
for either a 0.92 or 2.22~hr (or 4.44~hr for ellipsoidal
variations) periodicity. The first of these is favored by the
X-ray spectral properties discussed above.
We find no evidence for any 2--10~keV periodic modulation
at any of the above periods with 
2$\sigma$ upper limits to the semi-amplitude of any sinusoids 
of $\approxlt$1.5\%. If the optical variations
are indeed periodic, then the non-detection of an 
X-ray modulation is surprising. 
It is interesting to speculate on the origin of the partial covering. 
One explanation for the 
absorbed spectral component is that it is due to X-rays that pass
through the outer layers of an accretion disk. This
implies that the source is viewed close to the orbital
plane and would probably result in an X-ray orbital modulation,
which is not seen. Alternatively, the lack of an X-ray
orbital modulation may indicate that the absorbing material is located in an
optically thick accretion disk corona surrounding the
neutron star, or in a circum-binary disk or shell, perhaps
resulting from the accumulation of outflowing material.

\begin{acknowledgements}
The \sax\ satellite is a joint Italian-Dutch programme. 
L.~Sidoli acknowledges an ESA Fellowship. 
\end{acknowledgements}

\end{document}